\documentclass[%
 aip,
 amsmath,amssymb,
preprint,%
]{revtex4-2}

\usepackage{graphicx}
\usepackage{dcolumn}
\usepackage{bm}

\usepackage[utf8]{inputenc}
\usepackage[T1]{fontenc}
\usepackage{mathptmx}
\usepackage{etoolbox}
\usepackage{braket}
\usepackage{siunitx}

\makeatletter
\def\@email#1#2{%
 \endgroup
 \patchcmd{\titleblock@produce}
  {\frontmatter@RRAPformat}
  {\frontmatter@RRAPformat{\produce@RRAP{*#1\href{mailto:#2}{#2}}}\frontmatter@RRAPformat}
  {}{}
}%
\makeatother
\begin{document}

\preprint{AIP/123-QED}

\title[Quantum computing using floating electrons on cryogenic substrates]{Quantum computing using floating electrons on cryogenic substrates:  Potential And Challenges}
\author{A. Jennings}
\affiliation{RIKEN Center for Quantum Computing, 2-1 Hirosawa, Wako, Saitama, 351-0198, Japan}
\author{X. Zhou}%
\affiliation{RIKEN Center for Quantum Computing, 2-1 Hirosawa, Wako, Saitama, 351-0198, Japan}
\author{I. Grytsenko}
\affiliation{RIKEN Center for Quantum Computing, 2-1 Hirosawa, Wako, Saitama, 351-0198, Japan}
\author{E. Kawakami}
\email[email:]{A. Jennings: asher.jennings@riken.jp, E. Kawakami: e2006k@gmail.com}
 \homepage{https://sites.google.com/view/febqi/}
\affiliation{RIKEN Center for Quantum Computing, 2-1 Hirosawa, Wako, Saitama, 351-0198, Japan}
\affiliation{RIKEN Cluster for Pioneering Research, 2-1 Hirosawa, Wako, Saitama, 351-0198, Japan.}

\date{\today}

\begin{abstract}
In this review, we introduce a developing qubit platform: floating-electron-based qubits. Electrons floating in a vacuum above the surface of liquid helium or solid neon emerge as promising candidates for qubits, especially due to their expected long coherence times. Despite being in the early stages, a variety of recent experiments from different groups have shown substantial potential in this role. We survey a range of theoretical proposals and recent experiments, primarily focusing on the use of the spin state as the qubit state, wherein the spin and charge states are hybridized. Throughout these proposals and experiments, the charge state is coupled to an LC resonator, which facilitates both the control and readout mechanisms for the spin state via an artificially introduced spin-charge coupling.
\end{abstract}

\maketitle

\section{\label{sec:intro} Introduction}

To fully realize the promise of fault-tolerant quantum computing, it is imperative to have qubit systems that are scalable, have high fidelity, and are reasonably easy to manufacture. Here, we introduce the floating-electron-based (FEB) qubit. Essentially, the electrons are held in a vacuum above the cryogenic substrate, leading to long coherence times especially for the spin states. The electrons can be manipulated using electrodes placed underneath the cryogenic substrate. A single electron can serve as a qubit, and many qubits can be accommodated on a single chip, allowing a path to scalability and compatibility by applying semiconductor technology.

It was first theoretically predicted that a two-dimensional electron system (2DES) could be formed on the surface of some sort of cryogenic substrates in 1969~\cite{Cole1969-my,shikin1970motion}. To date, 2DES has been experimentally observed on liquid helium-3, liquid helium-4, solid neon, solid hydrogen, and solid deuterium~\cite{Crandall1972-hj,Kajita1985-tz,Kajita1983-ri,Monarkha2004Two-DimensionalSystems,Zavyalov-em}. Above the surface of those cryogenic substrates, a vacuum exists at low enough temperatures. In this manuscript, we focus on electrons on liquid helium-4 and solid neon in terms of their application to quantum computation.

It is instructive to compare the FEB qubits with commonly researched qubits. While any spin state even for an ensemble of electrons in any FEB system has not yet been observed, theoretical predictions suggest a spin dephasing time of over 100~s for an electron on helium~\cite{Lyon2006} surpasses any experimentally reported dephasing time in the solid state. 

Superconducting qubits, despite being the front-runners, grapple with challenges due to their size. The FEB system potentially allows for denser qubit alignment, accommodating more qubits within a cryogenic refrigerator. FEB qubits, trapped ions, and cold atoms share the fact that qubits exist in a vacuum and thus long coherence times are expected. However, the absence of laser requirements in FEB qubits might offer a distinct advantage for long-term scalability. In addition, an electron is much lighter than an ion or an atom and thus it is easier to move qubits on demand~\cite{Bradbury2011-ec,Sabouret2008-yj}. Compared to semiconductor quantum dots, the operational similarity of the proposed FEB qubits to semiconductor spin qubits is noteworthy as we will see in the later sections. Moreover, FEB qubits have the advantage of less demanding fabrication requirements (see also Sec.~\ref{sec:scalability_two-qubit_gate}). Furthermore, the integration of floating electrons with superconducting resonators, as highlighted in Refs.~\onlinecite{Schuster2010,Koolstra2019-mq,Zhou2022-nk,zhou2023electron}, shows potential for hybrid systems~\cite{Clerk2020-qx}.

Certainly, there are obstacles to overcome for the FEB qubits. Nearly a quarter-century has passed since electrons on helium were first proposed as a potential qubit platform~\cite{Platzman1999}. Although the important milestone of trapping a single electron on liquid helium was achieved~\cite{Papageorgiou2005,Glasson2005-sw,Rousseau2007}, an experimental demonstration of any qubit on liquid helium remains pending. The recent achievement of a charge qubit on solid neon marked a notable breakthrough~\cite{Zhou2022-nk, zhou2023electron}. Despite this being the sole qubit reported in the FEB system, this could herald a resurgence of interest in FEB qubits. The slow pace of advancement might have been mainly due to the challenges associated with accessing floating electrons. Reading out the quantum states of these electrons is no simple task. In contrast to semiconductor quantum dots, these electrons do not permit galvanic contact, thereby precluding measurement via electrical transport. However, recent research suggests that the most promising approach is to couple the electric dipole of an electron to an LC resonator, which is the approach adopted for the neon charge qubit. This methodology is discussed in detail in Sec.~\ref{sec:charge-photon}.

\section{Floating Electron Systems \label{sec:foating_syste}}

\begin{figure}
    \centering
    \includegraphics{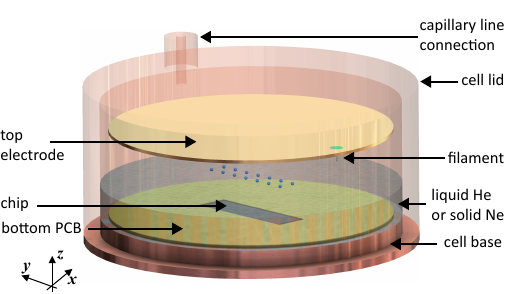}
    \caption{\label{fig:mock_cell}A basic floating-based electron experimental cell (typically 6~cm diameter and 5~cm height). The cell comprises two pieces, the lid and the base. On the base, a printed-circuit-board (PCB) can be mounted, with a cut for the nanofabricated chip (typically 2~mm by 7~mm). The lid has a hole and connection for a capillary line from which liquid helium or neon is filled. The whole cell is grounded. Above the PCB is a top electrode, which is also grounded. A cut in the top electrode is used to house a tungsten filament, from which electrons are deposited (see Appendix~\ref{sec:Appendix_deposition_electrons}). After electron deposition, positive voltages can be applied to electrodes on the chip to attract electrons.}
\end{figure}

\begin{figure}
\centering
\includegraphics[scale=1]{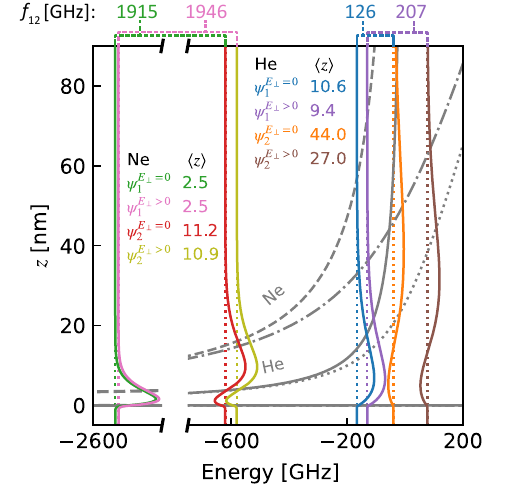}
\caption{\label{fig:states_transitions} The numerically solved eigenenergies (colored dotted lines), wavefunctions $\psi_n$ (colored solid lines), first-excited transition frequencies $f_{12}$ and average position $\langle z \rangle_n$ (legend) from the cryogenic substrate surface ($n=1$ for the Rydberg-ground state; $n=2$ for the Rydberg-1st-excited state). The solid (dashed) and dotted (dot-dashed) gray curves are the potentials of an electron on helium (neon) for $E_\perp=0$ and $E_\perp=\SI{15}{\kilo\volt\per\meter}$ (which is referred to as $E_\perp>0$ in the figure for clarity), respectively. The helium potentials were not plotted below \SI{-800}{\giga\hertz} for clarity. }
\end{figure}


An electron above the surface of liquid helium or solid neon induces a positive charge on the surface of the cryogenic substrate due to the relative permittivity difference between the cryogenic substrate and vacuum. This phenomenon can be captured by a positive image charge $\Lambda e$ placed in the cryogenic substrate at the same distance from the surface, where $ \Lambda=\frac{(\epsilon_r - 1)}{ (\epsilon_r + 1)}$ and $\epsilon_r$ is the relative permittivity of the cryogenic substrate (1.056 for liquid helium-4 and 1.244 for solid neon~\cite{Monarkha2004Two-DimensionalSystems,Jin2020-ym}). Thus, the electron is attracted toward the cryogenic substrate. However, due to the negative electron affinity of the cryogenic substrate, the electron generally does not penetrate into it, facing a potential barrier $V_0$ ($\approx1$~eV for liquid helium-4, and $\approx0.7$~eV for solid neon~\cite{Monarkha2004Two-DimensionalSystems,Jin2020-ym}) at the surface. Thus, the electron stays floating above the cryogenic substrate (see Fig~\ref{fig:mock_cell}). The total electric potential is then
\begin{equation}
    V(z) =
    \begin{cases}
    V_0, & \text{if}\ z\leq 0 \\
    - \frac{e^2 }{4 \pi \epsilon_0}\frac{\Lambda}{4  }\frac{1}{(z+z_0)}, & \text{if} \ z>0.
    \end{cases}
    \label{eq:potential}
\end{equation}
\noindent Here, $\epsilon_0$ is the vacuum permittivity and $e$ is the elementary charge. $z_0$ is a constant value introduced to be consistent with the spectroscopy measurement ($z_0 = 0.1$~nm for liquid helium-4 and = 0.23~nm for solid neon~\cite{Monarkha2004Two-DimensionalSystems,Jin2020-ym}).

 To solve the Schr\"{o}dinger equation with this potential analytically, one can simplify by setting $V_0 = \infty$ and $z_0 = 0$. Then the energy levels are given by
\begin{equation}
    E_{n_z} =  -  R_\infty \left(\frac{\Lambda}{4 }\right)^2  \frac{1}{ n_z^2},
\end{equation}
\noindent where $R_\infty=\frac{m_e e^4}{8 \epsilon_0^2 h^2}$ is the Rydberg constant, $m_e$ is the electron mass, $h$ is the Planck constant, and $n_z$ is the quantum number describing the direction perpendicular to the surface. These quantized eigenstates are usually referred to as the Rydberg states since it has a hydrogen-like energy spectrum~\cite{Monarkha2004Two-DimensionalSystems}\textsuperscript{,}~\footnote{In atomic physics, a Rydberg state refers to a state of an atom or molecule in which one of the electrons occupies a high principal quantum number orbital. However, in the field of electrons on helium, the ground state is also often called a Rydberg state.}. Alternatively, the Schr\"{o}dinger equation can be solved numerically (see Appendix~\ref{sec:Appendix_simulation}). The results are illustrated in Fig.~\ref{fig:states_transitions}, which are qualitatively consistent with the analytical results. The average position of the Rydberg-ground state is $\langle z \rangle_1=$ 10.6~nm, and 2.5~nm for helium-4, and solid neon, respectively.  In experiments, an additional perpendicular electric field $E_\perp$ can be introduced by applying voltages to the electrodes underneath the cryogenic substrate, to ensure the electron remains bound. The electric field $E_\perp>0$ pulls an electron closer to the surface and increases the energy requirement for the Rydberg transitions, but this effect is much smaller for neon (Fig.~\ref{fig:states_transitions}). 

\section{qubit state\label{sec:qubit_state}}

There are two choices for qubit states: the charge state or the spin state. Within the charge state, there are two types: the Rydberg state and the orbital state. As we have seen in the previous section, the perpendicular orbital motion is called the Rydberg state and hereafter, we refer to the parallel orbital motion as the orbital state. The orbital state is quantized due to the electrostatic confinement parallel to the surface of the cryogenic substrate, which is created by the voltages applied to the electrodes immersed in the cryogenic substrate~\cite{Schuster2010,Koolstra2019-mq,Zhou2022-nk,zhou2023electron} and the image potential of the electrodes~\cite{Dykman2003,kawakami2023blueprint}.

The first theoretical proposals~\cite{Platzman1999,Dykman2003,Lea2000,Dykman2000-do} focused on using the Rydberg-ground and 1st-excited states as the qubit state. Experimentally, many spectroscopic studies on the Rydberg state of an ensemble of many electrons were performed~\cite{Grimes1974-fg,Grimes1976SpectroscopyHelium,Grimes1979-rv,Collin2017Temperature-dependentHelium,Konstantinov2007-tw,Isshiki2007-oa,Konstantinov2008-gg,Kawakami2019} but no coherent-control of the Rydberg state of a single electron or an ensemble of electrons is realized yet. The Rydberg dephasing rate was theoretically calculated~\cite{Dykman2003} to be 100~Hz for $T=10~$mK and 100~kHz for $T=100~$mK. The Rydberg longitudinal relaxation rate is controversial. In Ref.~\onlinecite{Dykman2003}, it was calculated to be 10~kHz. Later, it was calculated to be at least 1~MHz, determined by the spontaneous emission of short-wavelength capillary waves of the liquid helium (ripplons)~\cite{Monarkha2007-el,Monarkha2010DecayRipplons}. The real-time measurement of the Rydberg decay for an ensemble of electrons on helium also demonstrated it to be 1~MHz when $T<600~$mK~\cite{Kawakami2021}.

Afterward, it was pointed out that the spin states of the electrons on liquid helium are expected to have an extremely long dephasing time $>100~$s, and therefore were proposed to be used as qubit states~\cite{Lyon2006}. One significant challenge associated with the electron spin state is its extremely small magnetic moment, making the direct control and readout of the spin state of a single electron challenging. As first proposed in Ref.~\onlinecite{Schuster2010} for electrons on helium, a solution is to introduce a coupling between the spin state and the charge state. By doing this, it becomes possible to control or read out the spin state through the more accessible charge state. After this first proposal, several theoretical proposals~\cite{Zhang2012,Dykman2023-hx,kawakami2023blueprint} for spin qubits on helium based on this method came out. In this review, we also primarily focus on this approach, believing it to be the direction in which efforts should be focused. We introduce several proposals on how to introduce the spin-orbit coupling in more detail in Sec.~\ref{sec:spin-orbit}.

\section{scalability and two-qubit gate\label{sec:scalability_two-qubit_gate}}

Scalability is fundamentally linked to the realization of a two-qubit gate, as it depends on the method used to connect the qubits. One advantage of the FEB system is that electrons are light, allowing for easy movement. Electrons can be physically moved in ways akin to the charge-coupled device method~\cite{Bradbury2011-ec, Sabouret2008-yj} or by utilizing a surface acoustic wave~\cite{Byeon2021-ht}. This mobility imparts greater flexibility to the qubit architecture, allowing for the connection of separate qubit arrays through electron movement, a strategy that has parallels in ion trap proposals~\cite{Kielpinski2002, Monroe2014-bl} and semiconductors~\cite{baart2016single, fujita2017coherent, mills2019shuttling, noiri2022shuttling, Zwerver2023}. Given this, our subsequent discussion will delve into the methods of interconnecting localized electrons within a qubit array, specifically in the context of realizing a two-qubit gate.

With the spin-charge coupling, each electron's spin state at a given site is linked to its charge state. Electrons on different sites interact via their electric interactions, which can occur through Coulomb interaction or by coupling with a microwave (MW) photon. As a result, even though the electrons are on different sites, their spin states can indirectly couple through these electric interactions. This allows for a two-qubit gate for the spin states of electrons on different sites.

\begin{figure}
    \centering
    \includegraphics[]{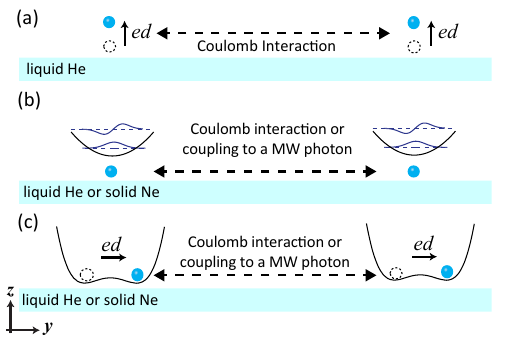}
    \caption{(a) The electric dipole moments along the $z$ axis induced by the transition from the Rydberg-ground state to the Rydberg-1st-excited state are coupled via Coulomb interaction. Blue circles and dotted circles represent the electrons in the Rydberg-1st-excited and in the Rydberg-ground state state, respectively. (b,c) The electric dipole moments along the $y$ axis are coupled by Coulomb interaction or by a MW photon through a superconducting resonator for electrons subjected to harmonic potentials in (b) and double well potentials in (c). (b) The blue lines show the wavefunctions of the ground and the 1st-excited orbital states of the electron.}
    \label{fig:two-qubit}
\end{figure}

The details of the spin-charge coupling at each electron's site are discussed in Sec.~\ref{sec:spin-orbit}. Different ways of realizing electric interaction are presented in Fig.~\ref{fig:two-qubit}. In Fig.~\ref{fig:two-qubit}(a), the transition between the Rydberg-ground state and the Rydberg-1st-excited state is involved. The electric dipoles $ed$, where $d=\langle z \rangle_2 -\langle z \rangle_1$, of distant electrons, are coupled via Coulomb interaction~\cite{Platzman1999,Dykman2003,Lea2000,kawakami2023blueprint}. In Fig.~\ref{fig:two-qubit}(b), the electric transition dipole moments of the orbital states are involved~\cite{Schuster2010,Dykman2023-hx,Zhang2012}. With each electron trapped within a harmonic confinement potential, the energy-level spacing is \( \hbar \omega_0 \), where $\hbar$ is reduced Planck constant, and the electric dipole is given by \( ed \),  where $d= l_0/\sqrt{2}$ denotes the electron's zero-point motion or the dot size, and $l_0=\sqrt{\hbar/m_e \omega_0}$ is referred to as the orbital spread. In Ref.~\onlinecite{Schuster2010}, the electric dipole moments of electrons couple to a MW photon, connecting distant electrons through superconducting resonators. In Refs.~\onlinecite{Dykman2023-hx, Zhang2012}, each electron is coupled to a normal mode of the collective in-plane vibration of the electrons, arising due to the Coulomb interaction. When the Coulomb interaction is used, typically electrons are separated by 1-3~$\mu$m~\cite{Dykman2023-hx,kawakami2023blueprint}. To enhance the electric dipole strength, a single dot created by a harmonic potential in Fig.~\ref{fig:two-qubit}(b) can be replaced with a double-quantum-dot (DQD) like structure, where an electron is subjected to a double-well potential~\cite{Burkard2020-gy,Harvey-Collard2022-jh}, as depicted in Fig.~\ref{fig:two-qubit}(c). With this approach, \( d \) now denotes the interdot distance.

\begin{figure}
    \centering
    \includegraphics[]{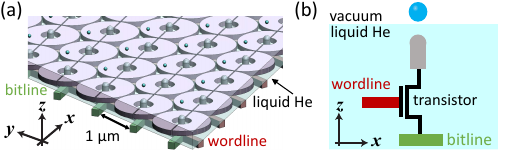}
    \caption{(a) Electrons (blue circles) are arranged in a two-dimensional array to facilitate a scalable quantum computer. The detailed geometry for a single site is illustrated in Fig.~\ref{fig:resonator}(b) and described in Ref.~\onlinecite{kawakami2023blueprint}. (b) At each intersection of the bitlines and wordlines, following a transistor, a central gray cylinder is positioned, atop which a single electron is floated.}
    \label{fig:scalabilty}
\end{figure}

Figure~\ref{fig:scalabilty} illustrates a method to scale up the number of qubits, based on the qubit design from Ref.~\onlinecite{kawakami2023blueprint}. Our goal is to integrate quantum bits at bitline and wordline intersections, echoing the DRAM method of classical computers. This approach was first introduced for semiconductor qubits~\cite{Veldhorst2017} that require a  proximity of approximately 100~nm between qubits to realize a two-qubit gate using the exchange interaction. With existing semiconductor technology (3~nm process, gate pitch=48~nm)~\cite{Moore2021}, one qubit requires an area of  $\approx 450 \times 450 $~nm$^2$ of this DRAM-based classical circuits~\cite{Veldhorst2017}. Thus, this method requires the incorporation of additional metallic gates that mediate electrostatic coupling over large distances~\cite{Trifunovic2012,Trifunovic2013} or further advances in semiconductor technology that are eagerly anticipated. Contrary to that, electrons on helium qubits benefit from a larger $\approx$1~$\mu$m spacing, thanks to the unscreened Coulomb interaction~\footnote{Unlike semiconductor qubits where electrons are located in a polarizable medium and the Coulomb interaction is reduced, here the Coulomb interaction retains its strength in a vacuum. This ensures that the strength of the electric dipole-dipole interaction is sufficient to enable a rapid two-qubit gate operation, even with electron spacing as wide as 1~$\mu$m. While the electrodes underneath the electrons may produce a screening effect, these do not significantly diminish the electric dipole-dipole interaction strength. In fact, as detailed in Appendix F of Ref.~\onlinecite{kawakami2023blueprint}, these interactions might be enhanced under certain conditions.}, promoting easier electrode positioning and scalability with existing semiconductor technology, even without the need for additional components.

\section{charge-photon coupling\label{sec:charge-photon}}

\begin{figure}
    \centering
    \includegraphics{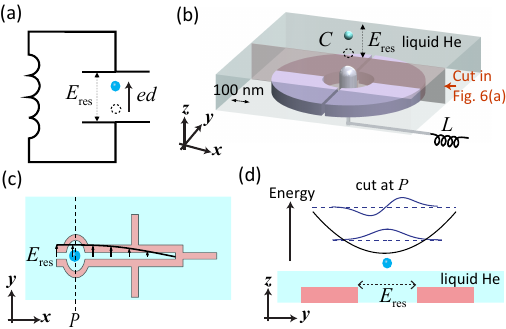}
    \caption{(a) Schematic of the coupling mechanism of an electron (light blue circle) to an LC resonator~\cite{Koolstra2019-ne}. The LC resonator is coupled to the external world capacitively or inductively (not shown). (b) Coupling of the Rydberg transition to the lumped-element LC resonator. Excerpted from Ref.~\onlinecite{kawakami2023blueprint}. The light blue sphere (dashed circle) represents an electron in the Rydberg-1st-excited (Rydberg-ground) state. The red plane shows the cut for the view in Fig.~\ref{fig:EDSR_mechanism}(a), where dimensions are given. (c,d) Coupling of the orbital state to the superconducting quarter wavelength resonator~\cite{Koolstra2019-mq,Koolstra2019-ne}. (c) Excerpted from Ref.~\onlinecite{Koolstra2019-ne}. (d) The pink rectangles in the right figure are parts of the resonator. The black line depicts a harmonic potential along the $y$ axis. The blue lines show the wavefunctions of the ground and the 1st-excited orbital states.} \label{fig:resonator}
\end{figure}

 Figure~\ref{fig:resonator}(a) shows a schematic representation of a floating electron coupling to an LC resonator. The electron experiences an electric field $E_\mathrm{res}$ in the capacitor and the electron forms an electric dipole $e d$. Assuming the electric field and the dipole direction align, the interaction between the resonator's microwave photon and the electric dipole is described by the Hamiltonian $H_\mathrm{int}= eE_\mathrm{res} d $.  Under the electric dipole and single-mode approximations, \( E_{\mathrm{res}} \) can be expressed as \( E_0(a + a^\dagger) \), where \( a \) and \( a^\dagger \) are the annihilation and creation operators for the resonator’s electromagnetic field mode, and \( E_0 \) represents the rms amplitude of the vacuum electric field. Consequently, this Hamiltonian simplifies to \( H_{\mathrm{int}} = \hbar g_c (a + a^\dagger) \), introducing the charge-photon coupling strength \( g_c \).  The expression for \( g_c \) is given by \( \hbar g_c = eE_0d = e \alpha V_0 \), where \( \alpha \) denotes the differential lever arm of the resonator electrode relative to the electron~\cite{Ibberson2021-uq}, $V_0$ is the rms voltage of the vacuum fluctuation. 

In Ref.~\onlinecite{kawakami2023blueprint}, it was proposed to use a lumped-element LC resonator to read out the transition between the Rydberg-ground state and the Rydberg-1st-excited state of a single electron on helium following the dispersive readout technique developed in semiconductor DQD~\cite{reilly2007fast,Gonzalez-Zalba2015,Vigneau2022-gj,Ibberson2021-uq} (see Appendix~\ref{sec:appendix_alternate} for other proposed read out methods). In this case, the electric dipole $ed$ couples to the $z$-electric field (Fig.~\ref{fig:resonator}(b)). With knowledge of the electrode geometry and the electron position, one can determine $\alpha$ using the Shockley-Ramo theorem~\cite{He2001,kawakami2023blueprint}. For the geometry proposed in Ref.~\onlinecite{kawakami2023blueprint}, $\alpha=0.01$ was calculated. In semiconductor qubits,  $\alpha=0.72$ was achieved~\cite{Ibberson2021-uq}; however, due to the geometry of the electrodes used for a floating electron in a vacuum, attaining a comparable $\alpha$ here is challenging. With this modest $\alpha$, together with a high-quality factor lumped LC resonator~\cite{Vigneau2022-gj,Ibberson2021-uq,Ahmed2018-he,Apostolidis2020-ck},  $g_c/2\pi \approx 0.1~$MHz can be achievable (see Appendix~\ref{sec:Appendix_LC}). Although reaching the charge-photon strong coupling regime is marginal given the Rydberg relaxation rate of 10~kHz-1~MHz for an electron on helium, it has been estimated that the above-mentioned  high-quality factor lumped LC resonator has high enough sensitivity to detect the Rydberg transition of a single electron in Ref.~\onlinecite{kawakami2023blueprint}. The transition between the Rydberg-ground state and the Rydberg-1st-excited state is induced by sending a resonant MW to the Rydberg transition energy ($f_\mathrm{12}\approx 127$~GHz) through a waveguide~\cite{Grimes1976SpectroscopyHelium,Collin2002MicrowaveHelium} and whether the transition occurs is read out by measuring the change in the reflected or transmitted 100~MHz signal from the LC resonator. Such an experiment for a single electron has not yet been achieved but is in progress. The principle experimental verification was conducted using many electrons on helium in a parallel plate capacitor~\cite{Kawakami2019}. The electric dipole induces the charge change to the parallel plate capacitor and this charge change is detected as a current. This detection was possible thanks to the $\approx 10^7$ electrons contributing to the signal although $\alpha=10^{-5}$ for one electron. Later, this method was also employed for $\approx 10^5$ electrons on thin helium film~\cite{Zou2022-fu}. Furthermore, by introducing the magnetic field gradient as proposed in Ref.~\onlinecite{kawakami2023blueprint}, the Rydberg transition energy becomes spin-dependent~\footnote{This is analogous to the Pauli spin blockade in semiconductor quantum dots where the electron tunneling between the DQD is dependent on the spin state~\cite{Ono2002} and using the aforementioned dispersive method on this blockade enabled spin state readout~\cite{Oakes2023-ys,Ciriano-Tejel2021-sz,Crippa2019-ir}.}. Consequently, the spin state can be read out by measuring the change in the reflected or transmitted signal from the LC resonator in a quantum-non-demolition manner~\cite{kawakami2023blueprint}.

In Ref.\onlinecite{Koolstra2019-mq}, a single electron was trapped at the electric field maxima of a quarter-wavelength superconducting resonator (Fig.~\ref{fig:resonator}(c)). As depicted in Fig.~\ref{fig:resonator}(d), the $y$ electric field couples to the $y$ direction orbital state. We assume a harmonic confinement potential along the $y$ axis and that the orbital energy-level spacing due to this potential is $\hbar \omega_0$. The electric dipole is represented by $ed$ as discussed in Sec.~\ref{sec:scalability_two-qubit_gate}. Compared to the lumped-element circuit, the electrode geometry used here yields a slightly larger value~\cite{Koolstra2019-ne} of $\alpha=0.03$. By increasing the resonance frequency to $\omega_r/ 2 \pi \approx 6~$GHz, the zero-point voltage fluctuation across the capacitor $V_0=\sqrt{ \hbar \omega_r/2C}$ is enhanced. By tuning the voltage applied to an electrode located close to the electron, the orbital energy was set to resonance with the microwave photon energy $\omega_0=\omega_r$, and the charge-photon coupling strength $g_c/2\pi=4.8~$MHz was measured. The high-quality factor Nb superconducting resonator exhibits a low loss rate $\kappa/2\pi=0.5$MHz. Unfortunately, the charge linewidth was measured to be $\gamma_c/2\pi=77$~MHz, preventing the system from reaching the strong coupling regime $g_c>\gamma_c, \kappa$. The estimated charge decoherence rate from this linewidth $\gamma_c/2\pi=77~$MHz is $\approx 100$ times larger than the theoretically estimated value~\cite{Schuster2010}. This discrepancy is believed to originate from helium surface level fluctuations, which are induced by the mechanical vibration caused by the pulse tube of the cryogenic refrigerator. Mechanical vibrations caused by pulse tubes can be reduced by  commercial or home solutions, or simply avoided by using a wet refrigerator\cite{OLIVIERI201773,DADDABBO201856, SCHMORANZER2019102, UHLIG2023103649}.

Strong coupling between a microwave photon and the charge state of a single electron has been achieved for the first time in the FEB system with electrons on solid neon~\cite{Zhou2022-nk}. This milestone was reached using the same device previously used for electrons on helium in Ref.~\onlinecite{Koolstra2019-mq}. Strong coupling could be achieved using solid neon primarily due to its longer charge coherence time compared to liquid helium. Not only is the solid substrate unaffected by the mechanical vibration, but the absence of capillary waves on the substrate also suggests that the intrinsic charge coherence time could be longer. The charge-photon coupling, the charge linewidth, and the cavity decay rate were measured to be $g_c/2\pi=3.5$~MHz, $\gamma_c/2\pi=1.7~$MHz, and $\kappa/2\pi=0.4~$MHz, respectively. Consequently, the condition for strong charge-photon coupling $g_c>\gamma_c,\kappa$ was satisfied. Real-time measurement revealed the charge coherence time of a single electron on solid neon to be 50~ns in Ref.~\onlinecite{Zhou2022-nk}. Subsequent research reported an extension of this time to a remarkable $50~\mu$s at a charge sweet spot \cite{zhou2023electron}. Both the  $T_1$ and $T_2$ times are on this order, setting a world record for electron charge states. The underlying mechanism for this extended coherence time remains to be investigated.  With 40 ns of gate time, the average gate fidelity of 99.97\% was measured using randomized benchmarking. To drive the qubit, a MW is sent through the resonator in the dispersive regime, resulting in the majority of the MW power being reflected. A dedicated drive MW line can be introduced to reduce the gate time further, while avoiding any heating effects, to reach an even higher gate fidelity. In both the experiments presented in Ref.~\onlinecite{zhou2023electron,Zhou2022-nk}, the single-electron charge spectrum exhibits a quadratic curve, suggesting that the electron is subject to a double well potential. Notably, the sample design was not initially aimed at producing a double well potential. Thus, the observed quadratic characteristics could potentially be ascribed to the surface roughness of the solid neon. This surface roughness may pose challenges as one aims to increase the number of uniform qubits. Additionally, accurately positioning a single electron becomes challenging because of the unforeseen image potential it experiences. Improvements in surface roughness could be realized by refining the device design and the solid neon growth process~\cite{zhou2023electron,Zhou2022-nk}.

\section{\label{sec:spin-orbit}spin-orbit coupling}

As discussed in Sec.~\ref{sec:qubit_state}, to have access to the spin state it is essential to utilize the coupling between the spin state and the charge state. Since there is no intrinsic spin-charge coupling, an artificial one must be introduced. While the spin-Rydberg interaction has also been explored~\cite{kawakami2023blueprint}, we restrict our discussion here specifically to the spin-orbit coupling. On one hand, large coupling is desirable for fast readout and control. On the other hand, one should avoid introducing additional relaxation or dephasing on the spin state through the coupling with the orbital state. Therefore, careful design and optimization of the system are necessary. Generally, there are two methods employed to create the spin-orbit coupling: using a current-carrying superconducting wire~\cite{Schuster2010,Dykman2023-hx,Zhang2012} or local ferromagnets~\cite{kawakami2023blueprint}. An advantage of using a current-carrying superconducting wire is the possibility of turning the spin-orbit coupling on and off, which is beneficial for certain qubit operations~\cite{Zhang2012}. One concern is that the magnitude of the current that can be sent through a superconducting wire is limited by the critical current of the superconductor\cite{Dykman2003,Kawakami2013} and thus the amplitude of the magnetic field gradient is limited. By placing local ferromagnets close to the electron, one can achieve a higher magnetic field gradient. A downside of using ferromagnets is that the spin-orbit coupling always exists, which might make some gate operations difficult~\cite{kawakami2023blueprint}. 

In the following, we revisit the cases considered in Ref.~\onlinecite{kawakami2023blueprint} and Ref.~\onlinecite{Schuster2010} for electrons on helium. Additionally, we present a study on the case of electrons on neon.

\begin{figure}
    \centering
    \includegraphics[scale=0.9]{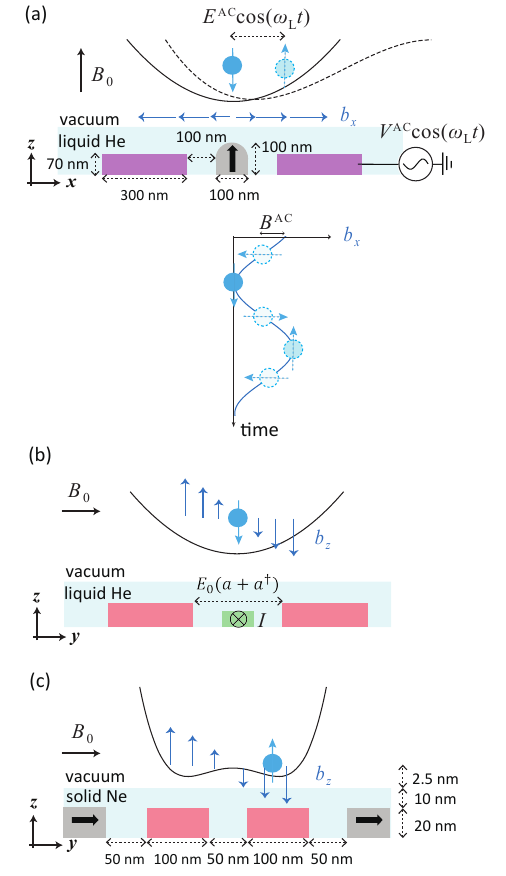}

\caption{\label{fig:EDSR_mechanism}Spin orbit coupling schemes for electrons (blue circles with an arrow depicting the direction of spin) floating above helium or neon. (a) The central pillar (dark gray) is made of a ferromagnetic material and surrounded by the two outer electrodes (purple -- see also Ref.~\onlinecite{kawakami2023blueprint} and Fig.~\ref{fig:resonator}(b)). An AC voltage $V^{AC}\cos(\omega_\mathrm{L} t)$ is applied to one of the outer electrodes, the electron experiences an AC electric field $E^{AC}\cos(\omega_\mathrm{L} t)$, cyclically modulating the position of the electron in the $x$ direction. When the ferromagnet is magnetized by an external field $B_0$ in the $z$ direction, the $x$ component of this field $b_x$ will have a gradient in the $x$ direction, shown by the blue arrows. Now the AC electric field becomes an effective AC magnetic field with amplitude $B^\mathrm{AC}$ for the electron. (b) A cut of a superconducting resonator (pink). The quantized electric field $E_0(a+a^\dagger)$ is shown by the arrows~\cite{Koolstra2019-mq,Zhou2022-nk}. A current-carrying central pin (green) is used to create a stray field $b_z$ (blue arrows) with a gradient in the $y$ axis~\cite{Schuster2010}. An external field $B_0$ is applied along the $y$ axis. (c) Also a cut of a superconducting resonator, but instead of the central strip, two ferromagnets (dark gray) are magnetized along the $y$ axis by $B_0$ to create $b_z$. The field gradient $\frac{\partial b_z}{\partial y}$ was calculated to be 0.1 mT/nm for cobalt~\cite{Pioro-Ladriere2008} sized 1.5~$\mu$m$\times$ 1.5~$\mu$m$\times$20~nm.}

\end{figure}

\subsection{Spin-state control for a single electron on helium}
\label{Spin-state control for a single electron on helium}

In the presence of an external magnetic field $B_0$ and the magnetic field gradient created by ferromagnets or a current-carrying superconducting wire, we employ electric dipole spin resonance (EDSR)~\cite{Nowack2007,Tokura2006,Pioro-Ladriere2008} to control the spin state (and thus to realize single-qubit gates), which works as follows. An AC voltage applied to one of the electrodes placed near the electron creates an AC electric field: $E^\mathrm{AC} \cos(\omega_\mathrm{L} t)$ along a direction, where $\omega_\mathrm{L}/2\pi\approx g\mu_B B_0/ h$ is the Larmor frequency, $g$ is the free electron Lande g factor, and $\mu_\mathrm{B}$ is the Bohr magneton. It modulates the electron's position approximately along the same direction as the AC electric field. In the presence of a magnetic field gradient along the direction of the electron's modulation—where the magnetic field gradient has components perpendicular to the external magnetic field—the electron experiences an effective AC magnetic field, inducing EDSR. Assuming that the confinement of the electron along the modulation direction can be approximated by a harmonic potential and that the energy-level spacing attributed to this potential is $\hbar \omega_0$, the amplitude of the effective AC magnetic field is calculated to be\cite{Schuster2010,Pioro-Ladriere2008b}  
\begin{equation} B^\mathrm{AC}=\Delta b \frac{eE^\mathrm{AC} l_0^2 \omega_0}{2\hbar (\omega_0^2- \omega_\mathrm{L}^2 )},
\label{eq_BAC}
\end{equation}
where $\Delta b$ is the gradient of the magnetic field perpendicular to the external magnetic field along the electron's modulation direction.  

In the case of the sample geometry used in Ref.~\onlinecite{kawakami2023blueprint} (Fig.~\ref{fig:EDSR_mechanism}(a)), an external magnetic field is applied normal to the surface (along the $z$ axis), and the ferromagnet is magnetized along the same axis, which creates a magnetic field gradient $\Delta b= \frac{\partial b_x}{\partial x} $ (see also Appendix~\ref{sec:Appendix_EDSR}). The electron's position is modulated parallel to the surface (along the $x$ axis) and experiences an external magnetic field $B^\mathrm{AC}$ along the $x$ axis, which results in EDSR. Ref.~\onlinecite{kawakami2023blueprint} calculated the Rabi frequency of the spin state induced by this EDSR mechanism is 100~MHz, the spin relaxation is reduced to 50~ms due to the spin-orbit coupling, and the spin coherence time stays intact.

\subsection{Spin-photon coupling for a single electron on helium}

\subsection{Spin-photon coupling for a single electron on neon}

Next, we consider how one can introduce the spin-orbit coupling for electrons on neon accommodating a superconducting resonator. In the charge-qubit experiments~\cite{zhou2023electron,Zhou2022-nk}, a single electron trapped on solid neon exhibits a spectrum that qualitatively resembles that of a semiconductor DQD spectrum, even if it was not intended, possibly due to the surface roughness of solid neon. In general, the charge-photon coupling (and thus the spin-photon coupling) is higher using a DQD than using a single-quantum-dot~\cite{Burkard2020-gy,Hu2012-hk}. Given these findings, we are now interested in how one can leverage this DQD-like feature for achieving spin-photon coupling. For a DQD, the spin-photon coupling is calculated as $g_s=g\mu_B B_0/\hbar$, where $B_0$ is given by
\begin{equation} B_0=\Delta b \frac{eE_0 d^2 }{4 (2t- \hbar \omega_\mathrm{L} )},
\label{eq_DQD_spin_photon}
\end{equation}
where $d$ is the interdot distance~\cite{Mi2018-co,Benito2017-ok,Samkharadze2018,Burkard2020-gy}. In the case of electrons on helium or neon, quantum tunneling between the dots is not expected~\cite{Dykman2023-hx}$^{\hspace{-0.1em},\hspace{-0.1em}}$\footnote{Due to the higher charging energy in a vacuum compared to in a semiconductor, owing to the lower dielectric constant in a vacuum, the electrochemical potential for an electron in a double-well potential tends to be set higher than the interdot barrier. Also note that the potential that an electron experiences on solid neon is described as a ring-shaped potential in Ref.~\onlinecite{Kanai2023-bo}. In both scenarios, quantum tunneling is not expected to occur.} and thus $t$ represents the orbital level spacing when the electron is located at the center between the two dots. The geometry shown in Fig. \ref{fig:EDSR_mechanism}(c) gives the magnetic field gradient at the position of the electron  $\Delta b= \frac{\partial b_z}{\partial y}=0.1~$mT/nm. Below we assume the interdot distance $d=100$~nm.
If $(2t- \hbar \omega_\mathrm{L})/h=1~$GHz, taking the charge-photon coupling $g_c/2\pi=e E_0 d/h= 3.5$~MHz based on Ref.~\onlinecite{Zhou2022-nk}, the spin-photon coupling is calculated to be $g_s/2\pi= 0.2~$MHz. The broadening attributed  to the nuclear spins for natural neon is expected to be~\cite{Chen2022-on} $\gamma_s/2\pi =10$~kHz. Due to the mixing between the spin state and the charge state, the spin decoherence rate is also influenced by the charge decoherence rate~\cite{Benito2017-ok}, given by $\left( \frac{g \mu_B \Delta b d }{ 2(2t- \hbar \omega_\mathrm{L} )} \right)^2 \gamma_c$ when $ g \mu_B \Delta b d/2  \ll 2t- \hbar \omega_\mathrm{L} $. Considering $\gamma_c/2\pi=0.36~$MHz based on Ref.~\onlinecite{zhou2023electron}, the spin decoherence rate attributed to the charge decoherence is calculated to be 7~kHz, which is the same order as the spin decoherence rate attributed to nuclear spins in solid neon~\cite{Chen2022-on}. Consequently, in this case, the hybridization between charge and spin does not significantly impact the spin coherence time. The strong spin-photon coupling $g_s>\gamma_s,\kappa$ is achieved with a low enough cavity decay rate.

If $(2t- \hbar \omega_\mathrm{L})/h=100~$MHz, the spin-photon coupling is calculated to be $g_s/2\pi= 2~$MHz. In this case, since $ g \mu_B \Delta b d/2  \approx  2t- \hbar \omega_\mathrm{L} $, the charge and spin states are hybridized to such an extent that the spin decoherence rate aligns with the charge decoherence rate, leading to $\gamma_s/2\pi \approx \gamma_c/2\pi =0.36~$MHz. The strong spin-photon coupling can be still achieved thanks to the long charge coherence time. However, the spin state's inherent strength, its extremely long coherence time, is lost. 

One can pursue enhancing the charge-photon coupling $g_c$ to attain a higher spin-photon coupling without compromising the spin coherence time. The vacuum electric field $E_0$ is increased by the voltage component of the vacuum fluctuations $  E_0 d = \alpha V_0$, where $V_0 \propto \sqrt{L_r/C_r}$  with the effective inductance $L_r$ and capacitance $C_r$ of the resonator~\cite{Beaudoin2016-os,Blais2004-ty}. By tailoring the sample geometry, one may enhance \(\alpha\).  Moreover, by maximizing the resonator's inductance and minimizing its capacitance, \(V_0\) can also be increased. A potential improvement is to employ NbTiN, in lieu of Nb used in Refs.~\onlinecite{Koolstra2019-mq,zhou2023electron,Zhou2022-nk} as the material for the superconducting resonator. NbTiN offers a high kinetic inductance and robustness against magnetic fields~\cite{Samkharadze2016-xh,Samkharadze2018}.

\section{\label{sec:conculsion} Outlook and Summary}

Realizing electron spin qubits in the FEB system is crucial, notably due to their expected long coherence times. In this review, we discussed controlling and detecting the spin state by hybridizing spin and charge states via artificially introduced spin-charge coupling and coupling the charge state to an LC resonator, considering this approach to hold significant promise. While the FEB system possesses pristine cleanliness and has advantages in scalability, addressing the challenges posed by the imperfectly smooth surface of solid neon and the susceptibility of superfluid helium to mechanical disturbances is essential to realize scalable quantum bits. Investigations are rigorously underway in both helium and neon platforms by various research teams worldwide.

\begin{acknowledgments}
We acknowledge Denis Konstantinov, David Rees, Dafei Jin, Fernando Gonzalez-Zalba, Rainer Dumke, Niyaz Beysengulov, and Atsushi Noguchi for useful discussions.

This work was supported by JST-FOREST (JPMJFR2039) and RIKEN-Hakubi program.
\end{acknowledgments}

\section*{Author Declarations}

\subsection*{Conflict of Interest}

The authors have no conflicts to disclose.

\subsection*{Author Contributions}

\noindent \textbf{Asher Jennings}: Software (equal); Validation (equal); Formal analysis (equal); Writing -- original draft (supporting); Writing -- review \& editing (equal); Visualization (equal).
\textbf{Xianjing Zhou}: Writing -- review \& editing (equal); Visualization (equal).
\textbf{Ivan Grytsenko}: Validation (equal); Investigation (equal); Data curation (equal).
\textbf{Erika Kawakami}: Conceptualization (equal); Methodology (equal); Validation (equal); Project Administration (equal); Writing -- original draft (lead); Writing -- review \& editing (equal); Visualization (equal); Supervision (equal); Funding acquisition (equal).

\section*{Data Availability}

The data that support the findings of this study are available from the corresponding authors upon reasonable request.

\section{Appendixes}

\appendix

\section{Deposition of electrons}\label{sec:Appendix_deposition_electrons}

\begin{figure}
\includegraphics[scale=1]{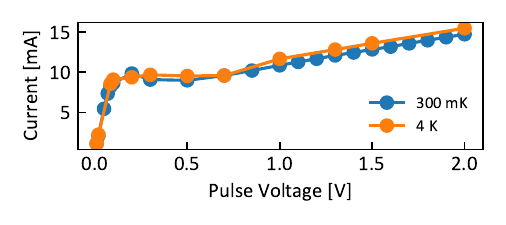}
\caption{\label{fig:filament} The current-voltage curve of a tungsten filament used to deposit electrons at temperatures of 4~K and 300~mK. The response is Ohmic at low voltages but then thermionic emission takes place at higher voltages (typically  $>\SI{0.2}{\volt}$) and shows a nonlinear behavior.}
\end{figure}

Electrons are deposited via the thermionic emission of electrons from a tungsten filament held a few millimeters above the cryogenic substrate. In experiments, people typically use 1.5V micro bulbs, model number CK1010-15 (CIR-KIT CONCEPTS, INC), which are fabricated for miniature dollhouses. The glass bulbs are cut to expose the filament. It is placed directly inside a cell where a vacuum is maintained. By sending a high enough voltage pulse (duration $\approx 3~$s), the filament heats up and thermionic emission of electrons occurs. We observe the electrons coming to the surface of the cryogenic substrate when the pulse voltage $>0.7~$V (Fig.~\ref{fig:filament}).

In the case of electrons on helium, it is easier to deposit electrons at $T> \SI{600}{\milli\kelvin}$, where there is still some helium vapor~\cite{Ando1978-pl,Saitoh1978-jc,Kawakami2021}. In this regime, thanks to the electron-helium vapor scattering, the kinetic energy of the electrons is transferred to the helium vapor, thus the electron will have lost momentum and more readily stay on the surface.

\section{Numerical Simulation for the Wavefunction}\label{sec:Appendix_simulation}

The Schr\"{o}dinger equation for a free electron on the surface of liquid helium or solid neon is given by

\begin{equation}
    \label{eq:APP_Schr\"{o}dinger}
    \left(-\frac{d^2}{dz^2} + \frac{2m}{\hbar^2}V(z)\right)\psi = \frac{2m}{\hbar^2}E\psi .
\end{equation}

\noindent The potential (see main text) is given by 

\begin{equation}
    V(z) =
    \begin{cases}
    V_0, & \text{if}\ z\leq 0 \\
    - \frac{e^2 }{4 \pi \epsilon_0}\frac{\Lambda}{4  }\frac{1}{(z+z_0)} + ezE_\perp, & \text{if} \ z>0.
    \end{cases}
    \label{eq:APP_potential}
\end{equation}

The Schr\"{o}dinger equation is solved by the finite difference method in a Python program. A potential is created from $z=\SI{-20}{\nano\meter}$ to $z=\SI{100}{\nano\meter}$, with steps of \SI{0.1}{\nano\meter}. A second-order, central finite difference matrix is created by FinDiff package~\cite{findiff}. After summing with the matrix created by the potential, the resulting eigenvalues and eigenfunctions are found with the SciPy package~\cite{2020SciPy-NMeth}. To retrieve the normalised wavefunctions from Fig.~1 in the main text, simply subtract the eigenenergy and divide by \SI{700}{\giga\hertz}.

\begin{figure}
    \centering
    \includegraphics[]{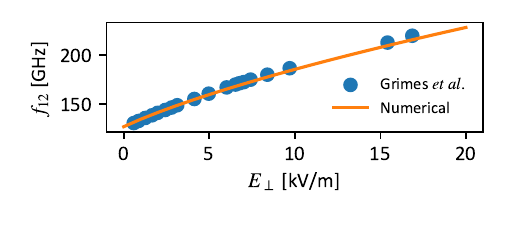}
    \caption{A comparison of the numerical calculation of the transition energy between the Rydberg-ground and Rydberg-1st-excited states to experimental data in Ref~\onlinecite{Grimes1976SpectroscopyHelium}.}
    \label{fig:grimes}
\end{figure}

The results are verified against experimental data from Ref. \onlinecite{Grimes1976SpectroscopyHelium} in Fig.~\ref{fig:grimes}. For solid neon, Ref. \onlinecite{Zavyalov-em} measured a higher $f_{12}$ than our simulation, as do other calculations\cite{Cole1969-my}.

\section{Detection of an escaped electron}
\label{sec:appendix_alternate}
\begin{figure}
    \centering
    \includegraphics[scale=1]{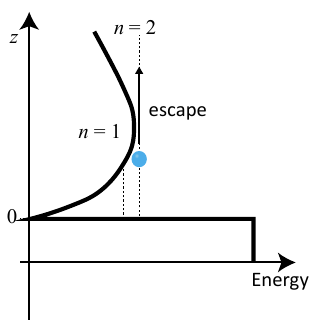}
    \caption{The electron-escape readout scheme described in the initial qubit proposal~\cite{Platzman1999}. Here, the perpendicular electric field $E_\perp < 0$ is tuned such that an electron in the Rydberg-ground state is still attracted to the surface, but an electron in the Rydberg-1st-excited state is no longer bound to the surface. The escaped electron can then be captured to perform a readout.}
    \label{fig:enter-label}
\end{figure}

While, in the main text, we discussed the readout of the qubit state by coupling the charge state to an LC resonator, the initial qubit proposal~\cite{Platzman1999} offers an alternative technique for electrons on helium. By applying a negative perpendicular electric field $E_\perp<0$, it is possible to create conditions in which the electron escapes from the surface when in the Rydberg excited states, and remains on the surface when in the Rydberg-ground state. Due to the lengthy electron loading process in this system, this readout strategy is not feasible during quantum computations. However, it may prove useful for end-of-calculation readouts. Potential methods to detect the escaped electron include using a superconducting micro-strip single-electron detector (SSED)~\cite{Shigefuji2023-wz} or an electron multiplier~\cite{Allen1947-pz}.

\section{Rydberg transition detection using an LC resonator}
\label{sec:Appendix_LC}
\subsection{Effective Hamiltonian}

In Eq.~(H1) of Ref. \onlinecite{kawakami2023blueprint},  the Hamiltonian of the first two-lowest Rydberg states of an electron integrated into an LC circuit and subjected to microwave (MW) excitation is given by
\begin{equation}
    H'=\frac{h f_\mathrm{0,Ry}}{2}s_z+eE^\mathrm{MW} z \cos(2\pi f_\mathrm{MW}t)+eE^m z\cos(2 \pi f_m t) \label{Hamiltonian_for_detection},
\end{equation}
where \( f_\mathrm{0,Ry} \) is the Rydberg transition frequency,  \( f_{\mathrm{MW}} \) is the MW frequency, $E^m$ is the electric field amplitude caused by the MW,  \( f_{m} \) is the frequency of the RF signal applied to the LC circuit (probe signal), and  \( E^m \) is the electric field amplitude caused by the probe signal.  $s_z=\ket{2}\bra{2}-\ket{1}\bra{1}$ and  $s_x=\ket{2}\bra{1}+\ket{1}\bra{2}$ are Pauli operators taking the Rydberg-ground state \( \ket{1}\) and the first-excited state \( \ket{2}\) as basis (see also Ref. \onlinecite{kawakami2023blueprint}).  In the rotating reference frame at frequency \(f_\mathrm{MW}\) and rotating wave approximation, the Hamiltonian can now be written as
\begin{equation}
H_{\mathrm{R}}= \frac{h \Delta f_{0} + e E^m d\cos(2 \pi f_{m} t)}{2} s_{z} + \frac{ e E^\mathrm{MW} z_{12}}{2}s_{x},
\label{eq:Hamiltonian_for_detection_R}
\end{equation}
where  \( \Delta f_{0} = f_{\mathrm{MW}} - f_\mathrm{0,Ry} \),  \(z_{12}=\bra{1}z\ket{2}\), and \( d = \langle z \rangle_{2} - \langle z \rangle_{1} \) denotes the distance between the Rydberg-ground state and the first-excited state as defined in the main text. For better comparison with semiconductor DQD~\cite{Petersson2012,Burkard2020-gy}, the Hamiltonian can be rewritten as
\begin{equation}
H_{\mathrm{R}}= \left( \frac{\epsilon}{2}+\frac{\delta \epsilon}{2} \cos(\omega_m t) \right) s_{z} + t s_{x},
\label{eq:Hamiltonian_same_as_semiconductor_qubits}
\end{equation}
where $\epsilon=h\Delta f_\mathrm{0,Ry}$, $\delta \epsilon=eE^md$, and $t=\frac{ e E^\mathrm{MW} z_{12}}{2}$. To facilitate our analysis, a transformation to a new Pauli basis is proposed, defined as follows:
\begin{align}
s_x &= R^\dagger \sigma_x R= \frac{2t}{\nu}\sigma_z +\frac{\epsilon}{\nu} \sigma_x, \nonumber \\
s_z &= R^\dagger \sigma_z R= \frac{\epsilon}{\nu}\sigma_z -\frac{2t}{\nu} \sigma_x,
\end{align}
where $R=\exp\left(-i\left( \frac{\pi}{2}-\theta \right)\frac{\hbar \sigma_y}{2}\right)$, $\theta=\arctan(\epsilon/2t)$, and $\nu=\sqrt{\epsilon^2+4t^2}$. By taking~\cite{Blais2004-ty,Peri2023-ii} $ \delta \epsilon\cos(\omega_m t)=\hbar g_c(a+a^\dagger)$, the Hamiltonian can be rewritten as 
\begin{equation}
H_{\mathrm{R}}= \frac{\nu}{2} \sigma_z  + \hbar g_c   (a+a^\dagger) \left(\frac{\epsilon}{\nu}\sigma_z -\frac{2t}{\nu} \sigma_x   \right)/2.
\end{equation}
The total Hamiltonian including the the LC resonator is 
\begin{equation}
    H_\mathrm{tot}=\hbar \omega_r a^\dagger a +H_\mathrm{R},
\end{equation}
where $\omega_r$ is the resonance frequency of the LC resonator. Following Ref.~\onlinecite{Petersson2012,Burkard2020-gy}, in the rotating frame with the unitary $U=\exp\left(-i\frac{\hbar \omega_m \sigma_z}{2}t-i\hbar \omega_m a^\dagger a t \right)$ and making a rotating wave approximation for a small probe signal ($\delta \epsilon \ll \nu$), an effective Hamiltonian is obtained as
\begin{equation}
H'_{\mathrm{tot}}= \hbar \Delta_0 a^\dagger a+ \frac{\hbar \Delta}{2} \sigma_z +g_\mathrm{eff}  (a \sigma_+ +a^\dagger \sigma_-)  ,
\end{equation}
where $\Delta_0=\omega_r-\omega_m$, $\Delta=\nu/\hbar -\omega_m$, $g_\mathrm{eff}=-\frac{t}{\nu}g_c$, and
$g_c=\frac{\delta \epsilon}{\hbar}=\frac{eE_0d}{\hbar}=\frac{e\alpha V_0}{\hbar}$ with $E_0$ and $V_0$ being the rms amplitude of the vacuum electric field and the rms voltage of the vacuum fluctuation, respectively. With a typical capacitance of a lumped-element LC resonator~\cite{Vigneau2022-gj,Ibberson2021-uq,Ahmed2018-he} $C=2~$pF and $\omega_r= 2 \pi \cdot 100~$MHz, the rms voltage of the vacuum fluctuation is calculated to be $V_0=\sqrt{\frac{\hbar \omega_r}{2C}}=130$~nV. Together with $\alpha=0.01$, the charge-photon coupling strength $g_c/2\pi=0.31~$MHz is calculated.

\subsection{Response from the LC resonator}

To investigate the signal response from the LC resonator caused by the Rydberg transition, we conduct the following analysis. For a transmission-style resonator, as an example, the transmission coefficient can be expressed as follows~\cite{Peri2023-ii,Ibberson2021-uq,Kohler2017-ly,Kohler2018-rj}:
 
\begin{equation}
    t_c=\frac{i \kappa}{ \Delta_0 +g_c^2 \chi -i \kappa/2}
    \label{eq:transmission_coeff}
\end{equation}
where $\kappa$ is the photon loss rate from the resonator's input and output ports and $\chi=\left(\frac{\hbar}{\delta \epsilon } \frac{\omega_m}{2 \pi} \int_0^{2\pi/\omega_m} dt e^{i\omega_m t}\langle s_z \rangle\right)^* $. It is worth noting that $ \langle \rho_{22}\rangle_I$ in Ref.~\onlinecite{kawakami2023blueprint} corresponds to $ \mathrm{Re} (\frac{\omega_m}{2 \pi} \int_0^{2\pi/\omega_m} dt e^{i\omega_m t}\langle s_z \rangle) $ and the change in capacitance $\Delta C= \frac{\Delta q\langle \rho_{22}\rangle_I}{u_m}$ corresponds to~\cite{Peri2023-ii,Petersson2010-de} $ (\alpha e)^2 \mathrm{Re}(\chi)/\hbar$, where $\Delta q =\alpha e$ and $\delta \epsilon=\alpha e u_m $. Furthermore, the signal-to-noise ratio (SNR) can be calculated, assuming photon noise is the predominant noise source, as follows~\cite{Blais2004-ty,Ibberson2021-uq}:
\begin{equation}
    \mathrm{SNR}=|\Delta t_c|^2 \frac{\bar{n}}{n_\mathrm{noise}} \frac{\kappa}{2\pi }t_\mathrm{int},
    \label{eq:SNR}
\end{equation}
where $\Delta t_c$ is the difference in the transmission coefficient (Eq.~\ref{eq:transmission_coeff}) with and without the Rydberg transition, $\bar{n}$ is the number of photons in the resonator, $n_\mathrm{noise}$ is the photon noise, and $t_\mathrm{int}$ is the integration time of the measurement. Using Eq.~\ref{eq:SNR}, the capacitance sensitivity~\cite{Brenning2006-wc,Ares2016-ll} is calculated to be
\begin{equation}
    S_c=\frac{\Delta C}{\sqrt{\mathrm{SNR}}}\sqrt{t_\mathrm{int}} =\frac{\Delta C}{|\Delta t_c |} \sqrt{\frac{n_\mathrm{noise}}{\bar{n} }\frac{2\pi}{\kappa}}.
\end{equation}
When considering the relaxation dynamics, it should be noted that equations developed for semiconductor DQD may not be directly applicable to the Rydberg states. Energy relaxation typically occurs from the high-energy eigenstate to the low-energy eigenstate. Hence, energy relaxation occurs from the high-energy eigenstate to the low-energy eigenstate of $\frac{\nu}{2} \sigma_z$ for semiconductor DQD, while it occurs from the high-energy eigenstate to the low-energy eigenstate of $\frac{h f_\mathrm{0,Ry}}{2} s_z$ for the Rydberg states. This key difference can be attributed to the need to be in the rotating frame of frequency $f_\mathrm{MW}$ to derive a Hamiltonian for the Rydberg states that is analogous to the semiconductor DQD (Eq.~\ref{eq:Hamiltonian_same_as_semiconductor_qubits}). Thus, although our prior analysis did not explicitly consider relaxation dynamics (and thus the analysis remains valid), applying some literature equations to the Rydberg states may necessitate certain modifications.

\section{Landau levels}\label{sec:Appendix_EDSR}
For the case considered in Fig.~5(a) in the main text and in Ref.~\onlinecite{kawakami2023blueprint}, differently from most of the cases in semiconductor quantum dots~\cite{Kawakami2014,Pioro-Ladriere2008} and other proposals for electrons on helium~\cite{Schuster2010,Dykman2023-hx,Zhang2012} where an external magnetic field is applied parallel to the 2DES, the external magnetic field is applied normal to the surface and thus, to more rigorously treat the system, one should consider landau levels.  It can be done by replacing $\omega_0$ and $\omega_\mathrm{L}$ in Eq.~(3) in the main text with $\tilde{\omega}=\omega_0 \sqrt{1+\omega_c^2/4 \omega_0^2}$ and $\omega_\mathrm{L}-\omega_c \approx \omega_\mathrm{L}/2$, respectively, where $\omega_c$ is the cyclotron frequency. Consequently, one obtains $B^\mathrm{AC}=\frac{\partial b_x}{\partial x} \frac{eE^\mathrm{AC} l_0^2 \tilde{\omega}}{2\hbar (\tilde{\omega}^2- \omega_\mathrm{L}^2 /4)}$ as the amplitude of the effective AC magnetic field. One advantage of applying an external magnetic field perpendicular to the 2DES is that it lifts the degeneracy of the system. Another way to lift the degeneracy is to make an elongated quantum dot as demonstrated in Ref.~\onlinecite{Koolstra2019-mq}.

\nocite{apsrev41Control}
\bibliography{library}
\end{document}